\documentclass[11pt,twoside]{article}

\usepackage{asp2014}

\aspSuppressVolSlug
\resetcounters

\begin{document}

\title{ShowerModel: A Python package for modelling cosmic-ray showers, their light production and their detection}

\author{Daniel Morcuende, and Jaime Rosado}
\affil{IPARCOS and Department of EMFTEL, Universidad Complutense de Madrid, E-28040 Madrid, Spain; \email{dmorcuen@ucm.es, jrosadov@ucm.es}}

\paperauthor{Daniel Morcuende}{dmorcuen@ucm.es}{0000-0001-9400-0922}{Universidad Complutense de Madrid}{IPARCOS and Department of EMFTEL}{Madrid}{Madrid}{E-28040}{Spain}
\paperauthor{Jaime Rosado}{jrosadov@ucm.es}{0000-0001-8208-9480}{Universidad Complutense de Madrid}{IPARCOS and Department of EMFTEL}{Madrid}{Madrid}{E-28040}{Spain}



\begin{abstract}
Cosmic-ray observatories necessarily rely on Monte Carlo simulations for their design, calibration and analysis of their data. Detailed simulations are very demanding computationally. We present a python-based package called \texttt{ShowerModel} to model cosmic-ray showers, their light production and their detection by an array of telescopes. It is based on parameterizations of both Cherenkov and fluorescence emission in cosmic-ray induced air showers. The package permits the modelling of fluorescence telescopes, imaging air Cherenkov telescopes, wide-angle Cherenkov detectors or any hybrid design.

\texttt{ShowerModel} was conceived as a tool to speed up calculations that do not require a full simulation or that may serve to complement complex Monte Carlo studies and data analyses (e.g., as a cross-check). It can also be used for educational purposes.
\end{abstract}
 
\section{Introduction}
Very-high-energy cosmic rays and gamma rays induce extensive air showers (EAS) when entering the atmosphere. Cherenkov and fluorescence light emitted by secondary charged particles is used as a proxy for studying the primary particles that initiate the particle cascades.

Design, calibration and data analysis of cosmic-ray and gamma-ray observatories strongly rely on Monte Carlo simulations of both the air shower and detector response. \texttt{CORSIKA} program \citep{corsika} is widely used for carrying out the first step of the simulation, whereas the second step depends on the detection technique. For example, in the case of imaging atmospheric Cherenkov telescopes (IACT), the program \texttt{sim\_telarray} \citep{Bernlohr2008} is commonly used. These detailed simulations are currently very demanding computationally.

We present a fast python package called \texttt{ShowerModel} to compute the light emission in air showers and its detection by an array of telescopes \citep{ShowerModel}. This tool can speed up calculations that do not require a full simulation or that may serve to complement complex Monte Carlo studies and data analyses (e.g., as a cross-check). It can also be used for educational purposes. A similar approach was presented previously in \citet{Vuillaume2017}.

Functionalities and examples of results obtained with \texttt{ShowerModel} are shown. Several possible remarkable applications of this software are also briefly discussed.

\section{ShowerModel}
The package comprised several functions and classes to model air-shower development and detection (see Figure \ref{block_diagram}). Both gamma and proton-like 1D air showers can be generated using analytical Greisen or Gaisser-Hillas formulas \citep{Greisen1956, GaisserHillas1977}. Simulation-generated longitudinal profiles can also be input. Different flat atmospheric models are available.

\articlefigure[width=.8\textwidth]{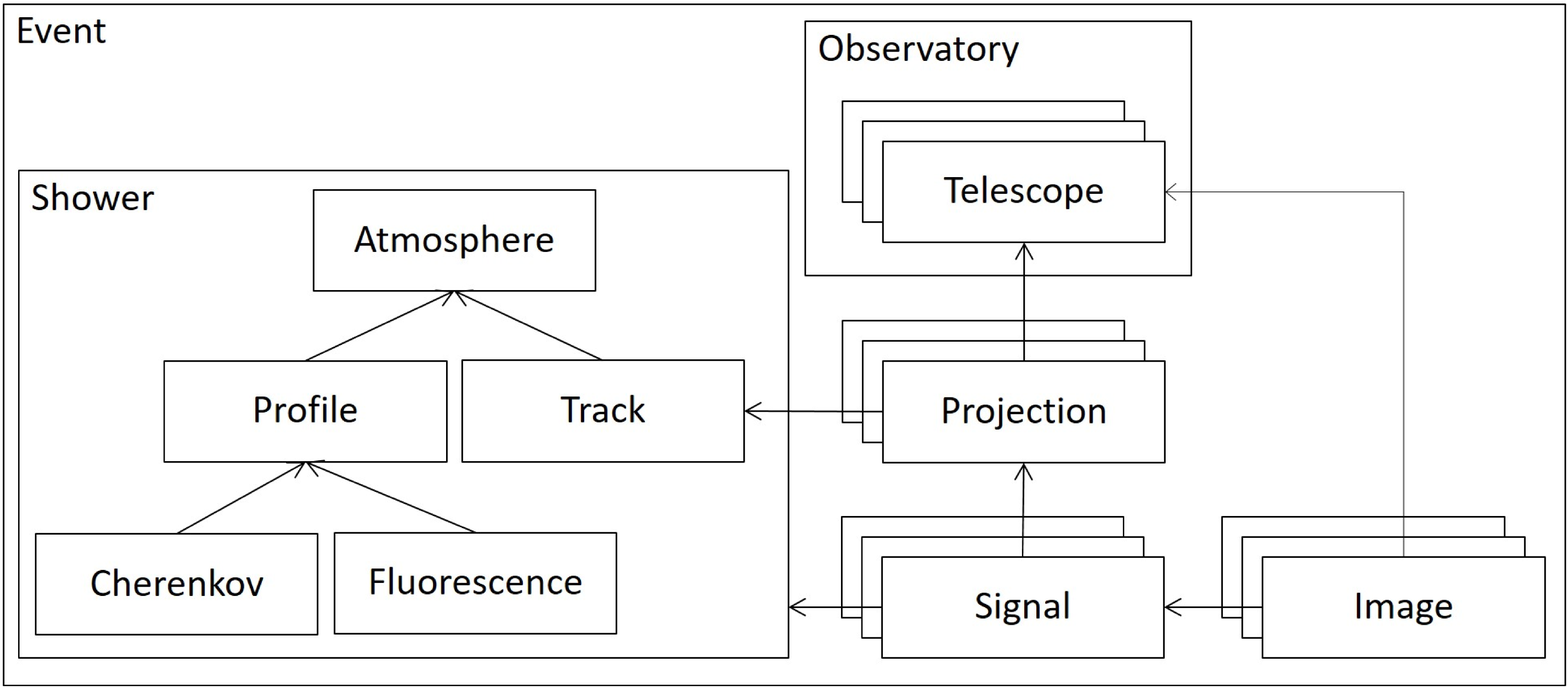}{block_diagram}{Class dependency of \texttt{ShowerModel}.}

The code uses detailed parameterizations of both fluorescence and Cherenkov light emission and angular distribution \citep{Morcuende2019, Nerling2006}. The time-varying light intensity reaching a telescope is readily computed from geometry. Rayleigh scattering losses are also included.

Telescope objects are highly configurable (e.g., quantum efficiency, pointing direction, field of view, detection area) to model fluorescence telescopes, imaging air Cherenkov telescopes, wide-angle Cherenkov detectors or hybrid designs. Camera images are produced using the Nishimura-Kamata-Greisen (NKG) function \citep{Nishimura1958, Greisen1956} to describe the lateral distribution of electrons in the air shower.

\section{Functionalities}

\texttt{ShowerModel} permits the calculation of:

\begin{itemize}
    \item Projection of shower tracks in local coordinates \texttt{alt}/\texttt{az} as well as the telescope field of view coordinates \texttt{theta}/\texttt{phi} (Figure \ref{fluorescence_telescope}, left).
    
    \articlefigure[width=.8\textwidth]{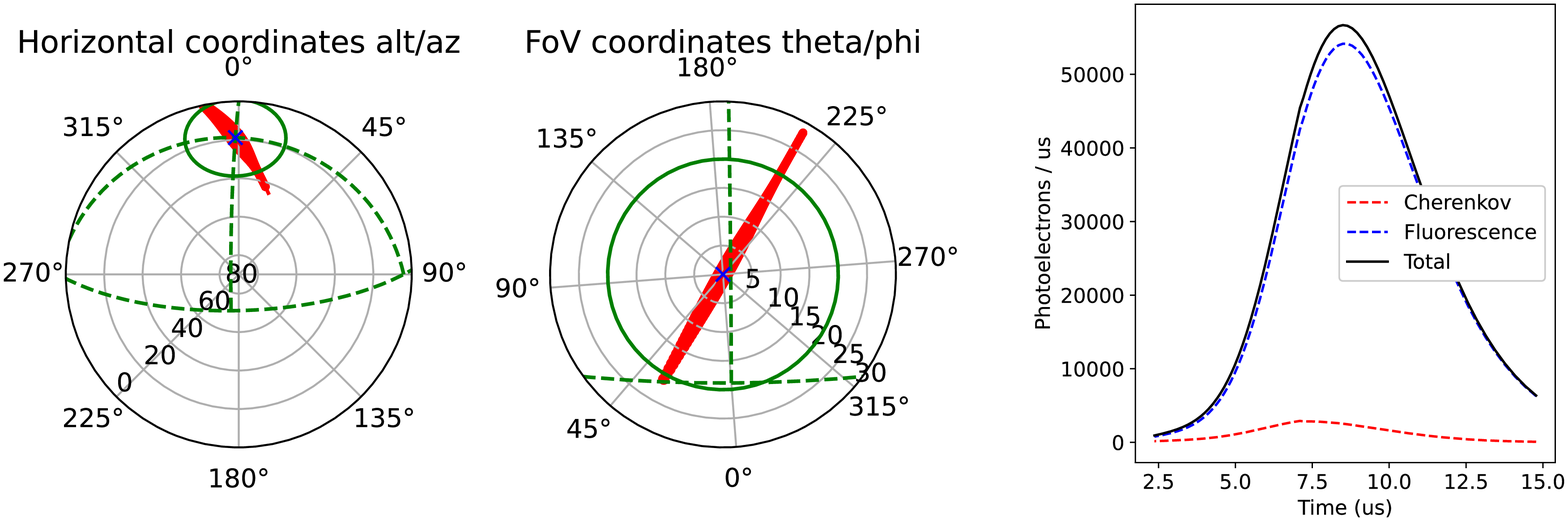}{fluorescence_telescope}{\emph{Left:} Shower track projections of an air shower observed by a fluorescence telescope.  \emph{Right:} Signal produced from the two light components.}
    
    \item Longitudinal shower profiles: energy deposit and light production.
    
    \item Cherenkov and fluorescence photon densities on ground (Figure \ref{2d_distr}).
    
    \articlefigure[width=0.8\textwidth]{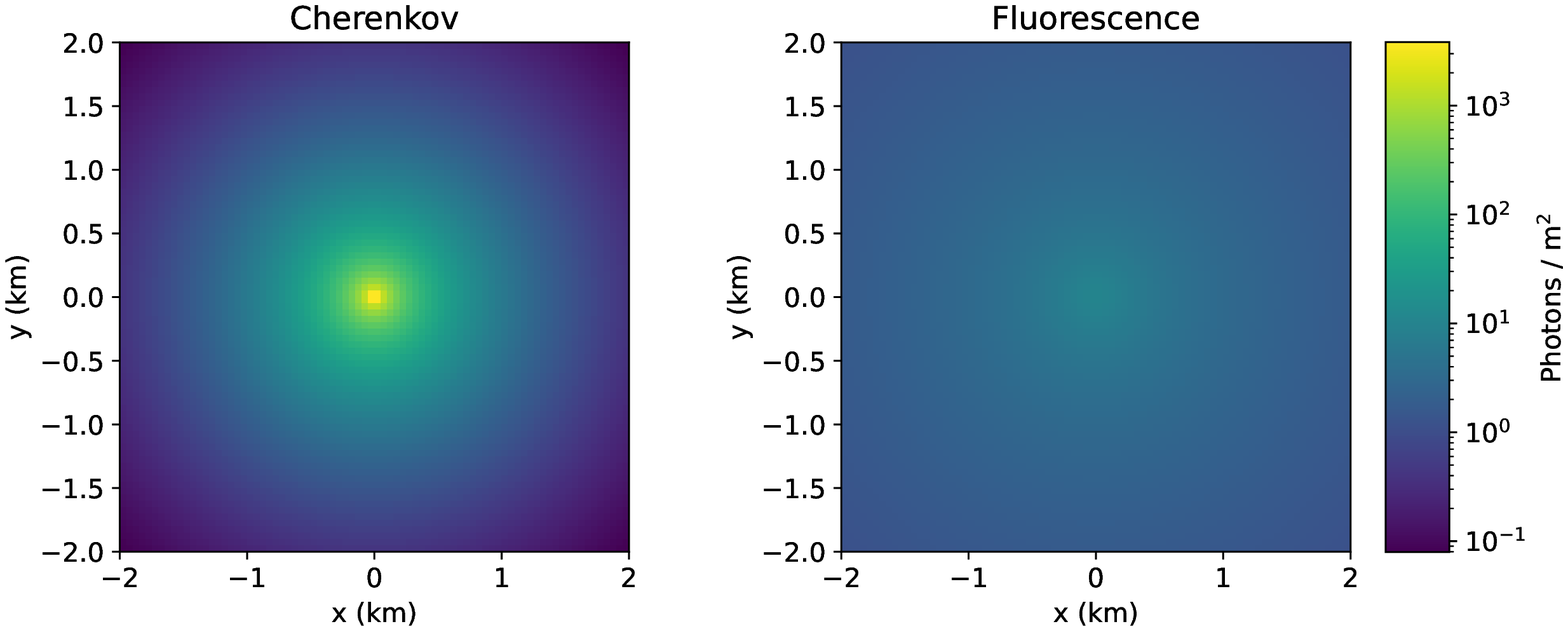}{2d_distr}{Cherenkov and fluorescence photon densities at ground level.}
    
    \item Time evolution of signal in a telescope (Figure \ref{fluorescence_telescope}, right).
    
    \item Shower events detected by an array of telescopes (Figure \ref{obs_event}).
    
    \articlefigure[width=.4\textwidth]{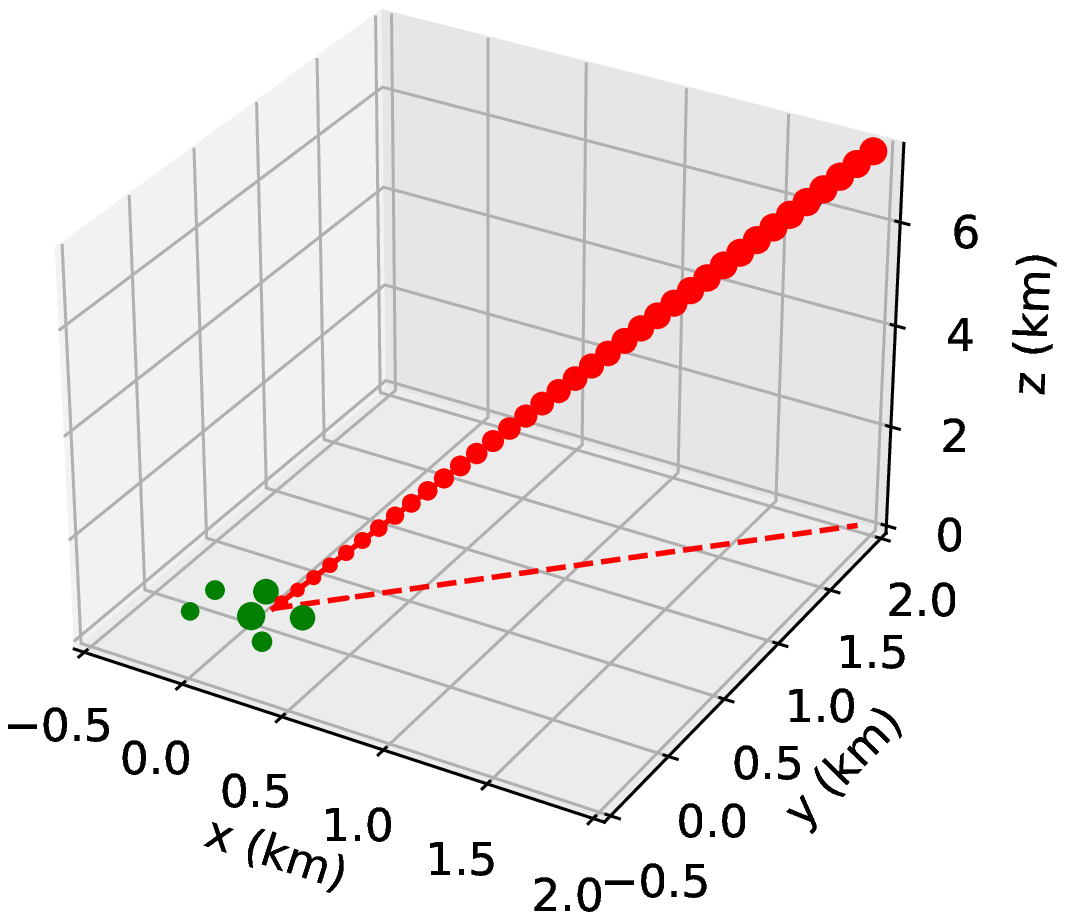}{obs_event}{Air shower observed by an observatory of 6 telescopes. Green dots mark the telescope positions and their relative signal sizes.}
    
    \item Camera images in customized telescopes (Figure \ref{camera_images}).

    \articlefigure[width=.8\textwidth]{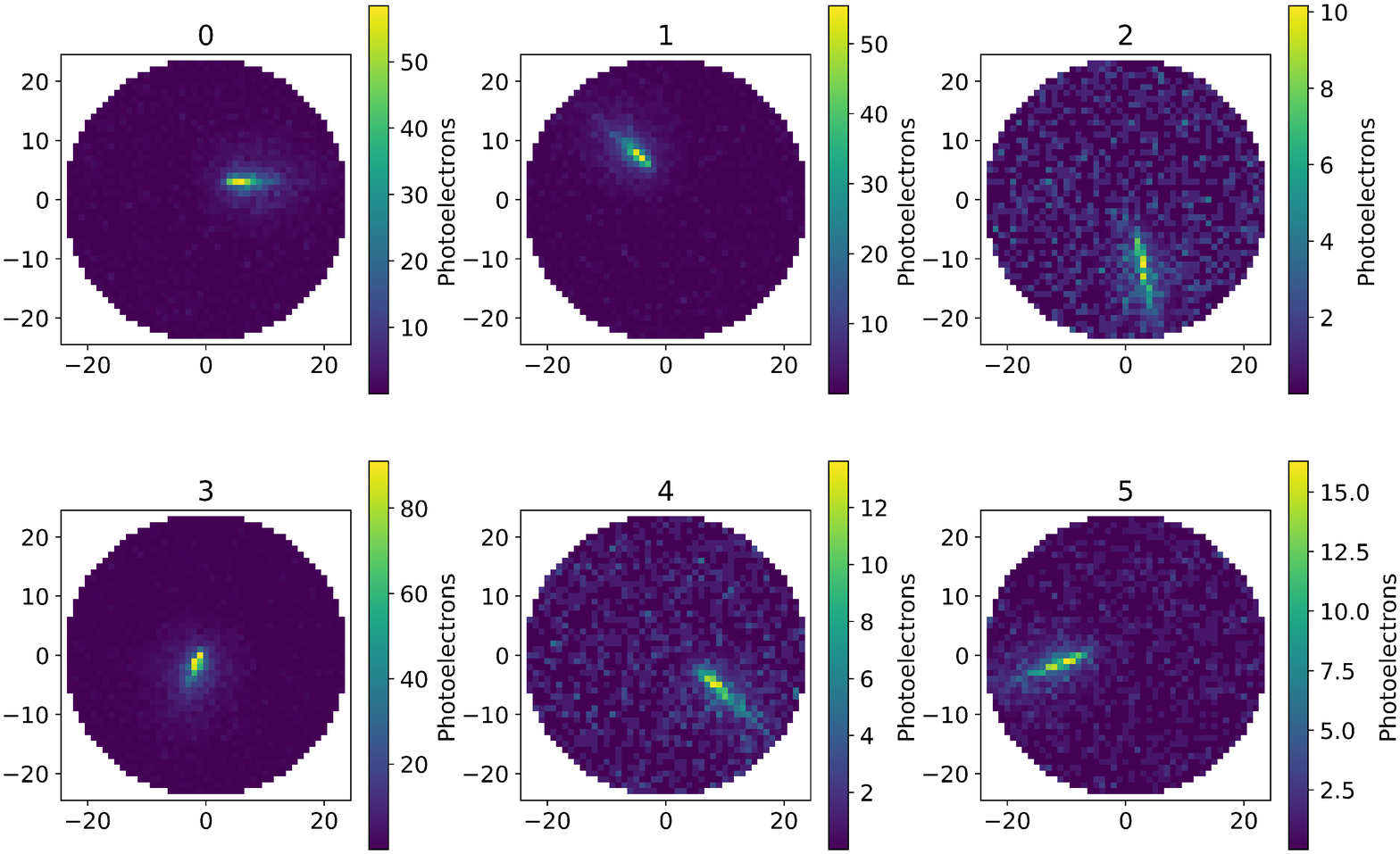}{camera_images}{Camera images of an air shower observed by an array of six IACTs.}
    
\end{itemize}

\section{Possible remarkable use cases}

The proposed tool can serve as a complement of the analysis pipeline of cosmic-ray observatories. For instance, it might be used to fast evaluate if changes in the observation conditions (e.g., atmospheric parameters and night sky background) make necessary new simulations or instrument response functions. In addition, it may be useful to explore new observatory configurations or detection techniques that exploit both Cherenkov and fluorescence signals (see \citet{ICRC2015} and \citet{Thesis_Sailer2020}).

\acknowledgements We gratefully acknowledge support from Spanish MINECO (contract FPA2017-82729-C6-3-R) and the European Commission (E.U. Grant Agreement 653477). D.~Morcuende acknowledges a predoctoral grant UCM-Harvard University (CT17/17-CT18/17) from Universidad Complutense de Madrid.

\bibliography{P4-176}  

\begin{thebibliography}{}
\expandafter\ifx\csname natexlab\endcsname\relax\def\natexlab#1{#1}\fi
\expandafter\ifx\csname url\endcsname\relax
  \def\url#1{\texttt{#1}}\fi
\expandafter\ifx\csname urlprefix\endcsname\relax\def\urlprefix{URL }\fi
\providecommand{\eprint}[2][]{\url{#2}}

\bibitem[{Bernl{\"{o}}hr(2008)}]{Bernlohr2008}
Bernl{\"{o}}hr, K. 2008, Astropart. Phys., 30, 149. \eprint{0808.2253},
  \urlprefix\url{https://doi.org/10.1016/j.astropartphys.2008.07.009}

\bibitem[{Contreras et~al.(2016)Contreras, Rosado, Arqueros, L{\'{o}}pez,
  Barrio, \& Nievas}]{ICRC2015}
Contreras, J.~L., Rosado, J., Arqueros, F., L{\'{o}}pez, M., Barrio, J.~A., \&
  Nievas, M. 2016, PoS, ICRC2015, 993.
  \urlprefix\url{https://doi.org/10.22323/1.236.0993}

\bibitem[{{Gaisser} \& {Hillas}(1977)}]{GaisserHillas1977}
{Gaisser}, T.~K., \& {Hillas}, A.~M. 1977, in International Cosmic Ray
  Conference, vol.~8 of International Cosmic Ray Conference, 353.
  \urlprefix\url{https://ui.adsabs.harvard.edu/abs/1977ICRC....8..353G}

\bibitem[{Greisen(1956)}]{Greisen1956}
Greisen, K. 1956, JG Wilson, Amsterdam, Netherlands, 3

\bibitem[{Heck et~al.(1998)Heck, Knapp, Capdevielle, Schatz, \&
  Thouw}]{corsika}
Heck, D., Knapp, J., Capdevielle, J.~N., Schatz, G., \& Thouw, T. 1998,
  Forschungszentrum Karlsruhe, FZKA 6019, 1

\bibitem[{Kamata \& Nishimura(1958)}]{Nishimura1958}
Kamata, K., \& Nishimura, J. 1958, Progress of Theoretical Physics Supplement,
  6, 93. \urlprefix\url{https://doi.org/10.1143/PTPS.6.93}

\bibitem[{Morcuende \& Rosado(2020)}]{ShowerModel}
Morcuende, D., \& Rosado, J. 2020, {JaimeRosado/ShowerModel v0.1.4}.
  \urlprefix\url{https://github.com/JaimeRosado/ShowerModel}

\bibitem[{Morcuende et~al.(2019)Morcuende, Rosado, Contreras, \&
  Arqueros}]{Morcuende2019}
Morcuende, D., Rosado, J., Contreras, J., \& Arqueros, F. 2019, Astroparticle
  Physics, 107, 26 .
  \urlprefix\url{https://doi.org/10.1016/j.astropartphys.2018.11.003}

\bibitem[{Nerling et~al.(2006)Nerling, Bl{\"{u}}mer, Engel, \&
  Risse}]{Nerling2006}
Nerling, F., Bl{\"{u}}mer, J., Engel, R., \& Risse, M. 2006, Astropart. Phys.,
  24, 421. \urlprefix\url{https://doi.org/10.1016/j.astropartphys.2005.09.002}

\bibitem[{Sailer(2020)}]{Thesis_Sailer2020}
Sailer, S. 2020, Ph.D. thesis, Heidelberg University, Combined Faculty of
  Natural Sciences and Mathematics, Germany.
  \urlprefix\url{http://www.ub.uni-heidelberg.de/archiv/29105}

\bibitem[{Vuillaume et~al.(2017)Vuillaume, Gat\'{e}, Maurin, Jacquemier, \&
  Lamanna}]{Vuillaume2017}
Vuillaume, T., Gat\'{e}, F., Maurin, G., Jacquemier, J., \& Lamanna, G. 2017,
  PoS, ICRC2017, 772. \urlprefix\url{https://doi.org/10.22323/1.301.0772}

\end{thebibliography}

\end{document}